\documentclass[12pt,twoside]{article}
\usepackage{fleqn,espcrc1}
\usepackage[dvips]{graphicx}

\newcommand{\xp}{\ensuremath{x_{I\!\!P}}}
\newcommand{\qq}{\ensuremath{Q^{2}}}
\newcommand{\FD}{\ensuremath{F_{2}^{D(3)}}}
\newcommand{\dcs}{\ensuremath{\sigma_{d}}}
\newcommand{\aem}{\ensuremath{\alpha_{em}}}

\newcommand{\D}{\ensuremath{\mbox{d}}}

\begin{document}
\begin{flushright}
CERN-TH/99-306 \\
MC-TH-99-14\\
October 1999\\
\end{flushright}
{\large  Predicting $F_{2}^{D(3)}$ from the dipole cross-section} \\

\noindent J.~R.~Forshaw\footnote{Theory Division, CERN,
1211 Geneva 23, Switzerland.},
G.~R.~Kerley
and G.~Shaw$^{a}$

\vspace*{0.5cm}

\noindent $^{a}$Theoretical Physics Group, Department of Physics and Astronomy,\\
The University of Manchester, M13 9PL, UK
\vspace*{0.5cm}

\begin{abstract}
We employ a parameterisation of the proton dipole cross section previously extracted from electroproduction and photoproduction data to predict the diffractive structure function $ \FD(\qq, \beta, \xp)$. 
Comparison with HERA H1 data yields good agreement.
\end{abstract}
\section{Introduction}
The proton dipole cross section \dcs\ is a universal quantity in singly dissociative diffractive $\gamma$ p processes.~\cite{bib:dcs_nik,bib:pom_DD_nik} 
It is simply the total cross section for scattering  a q\={q} pair of a given size and energy in the photonic fluctuation off the proton target.
We make use of a parameterisation used to extract \dcs\ from a fit to electroproduction and photoproduction $\gamma$ p total cross-section data~\cite{bib:our} to predict the diffractive structure function $ \FD(\qq, \beta, \xp)$.

\begin{figure}[htb]
    \includegraphics[width=10cm, height=3cm]{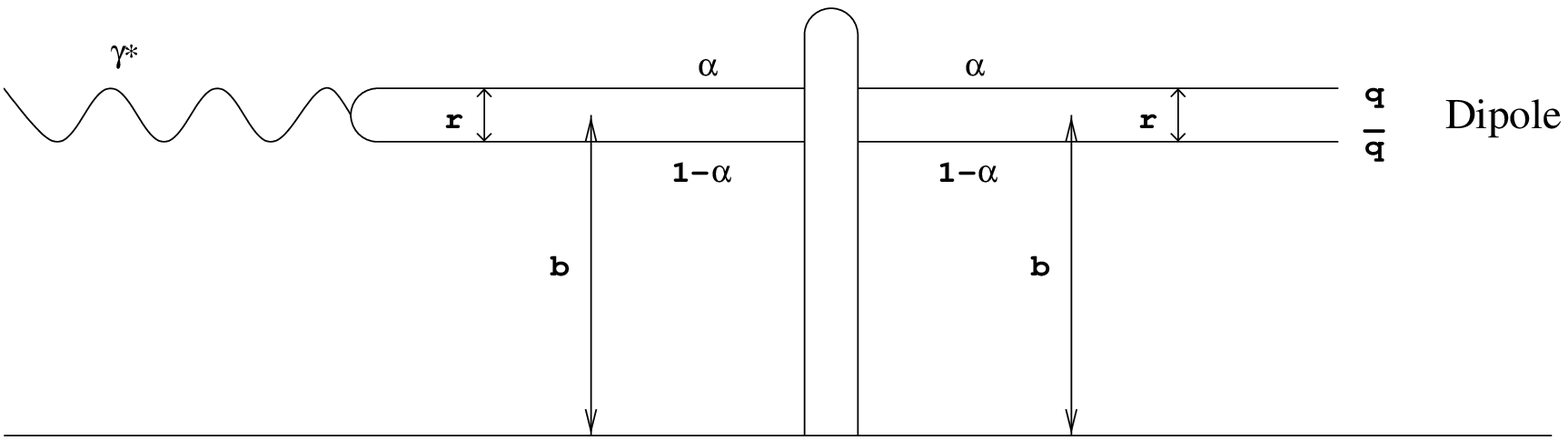}
    \label{fig:fluct}
    \caption{The dipole fluctuation of the incoming photon.}
\end{figure}
\section{Functional forms}

%
The dipole cross-section is in general a function of three variables (Figure~\ref{fig:fluct}):
  $s = W^2$, the CMS energy squared of the photon proton system;
  $r$, the transverse separation averaged over all orientations of the q\={q} pair; and
  $\alpha$, the fraction of the incoming photon light cone energy possessed by one member of the  q\={q} pair. 
We assumed a form with two terms, each  with a  Regge type $s$ dependence and no dependence on $\alpha$.
 Specifically, we assumed 
\begin{equation}
  \dcs(s,r)  =  a_{0}^{S}\left(1 - \frac{1}{1 + (a_{1}^{S}r + 
a_{2}^{S}r^{2})^{2}}\right)(r^{2}s)^{\lambda_{S}} + (a_{1}^{H}r + a_{2}^{H}r^{2} + a_{3}^{H}r^{3})^{2}
  \exp(-\nu_{H}^{2}r) (r^{2}s)^{\lambda_{H}}.  
\end{equation}
The dipole cross section is related to the photon proton cross section via 
\begin{equation}
\sigma^{L,T}_{\gamma^{*},p} = \int \mbox{d}\alpha \mbox{d}^{2}r \ |\psi_{\gamma}^
{L,T}(\alpha,r)|^{2} \dcs(s,r,\alpha)  
\end{equation} where \(\psi_{\gamma}^{L,T}(\alpha,r)\) are the longitudinal and transverse components of the 
light cone photon wave function.
For the photon wave function itself, we used the tree level QED expression~\cite{bib:dcs_nik,bib:wf} modified by a factor $f(r)$ to represent confinement effects:
\begin{eqnarray}
  \label{eq:psi^2}
  |\psi_{L}(\alpha,r)|^{2} & =  & \frac{6}{\pi^{2}}\aem\sum_{q=1}^{3}e_{q}^
{2}Q^{2}\alpha^{2}(1-\alpha)^{2} K_{0}^{2}(\epsilon r) \times f(r)\\
  |\psi_{T}(\alpha,r)|^{2} & = & \frac{3}{2 \pi^{2}}\aem\sum_{q=1}^{3}e_{q}^
{2} \left\{[\alpha^{2} + (1-\alpha)^{2}] \epsilon^{2} K_{1}^{2}(\epsilon r) + m_{f}^{2} 
K_{0}^{2}(\epsilon r) \right\} \times f(r)
\end{eqnarray}
and
\begin{equation}
  \label{eq:peak}
  f(r) = \frac{1 + B \exp(- c^{2} (r - R)^{2})}{1 + B \exp(- c^{2} R^{2})}.
\end{equation}
Here 
\(
 \epsilon^{2} = \alpha(1-\alpha)Q^{2} + m_{f}^{2}\; ,
\) 
$K_{0}$ and $K_{1}$ are modified Bessel functions and the sum is over 3 light quark flavours, with a generic mass of assumed value $m_f^2 = 0.08$ GeV$^2$.  
The values of the constants $B$, $c^{2}$, and $R$ were generated by the fit.
%
\section{Calculating $ \FD$}
\begin{figure}[htb]
  \includegraphics[height=6cm,width=6cm]{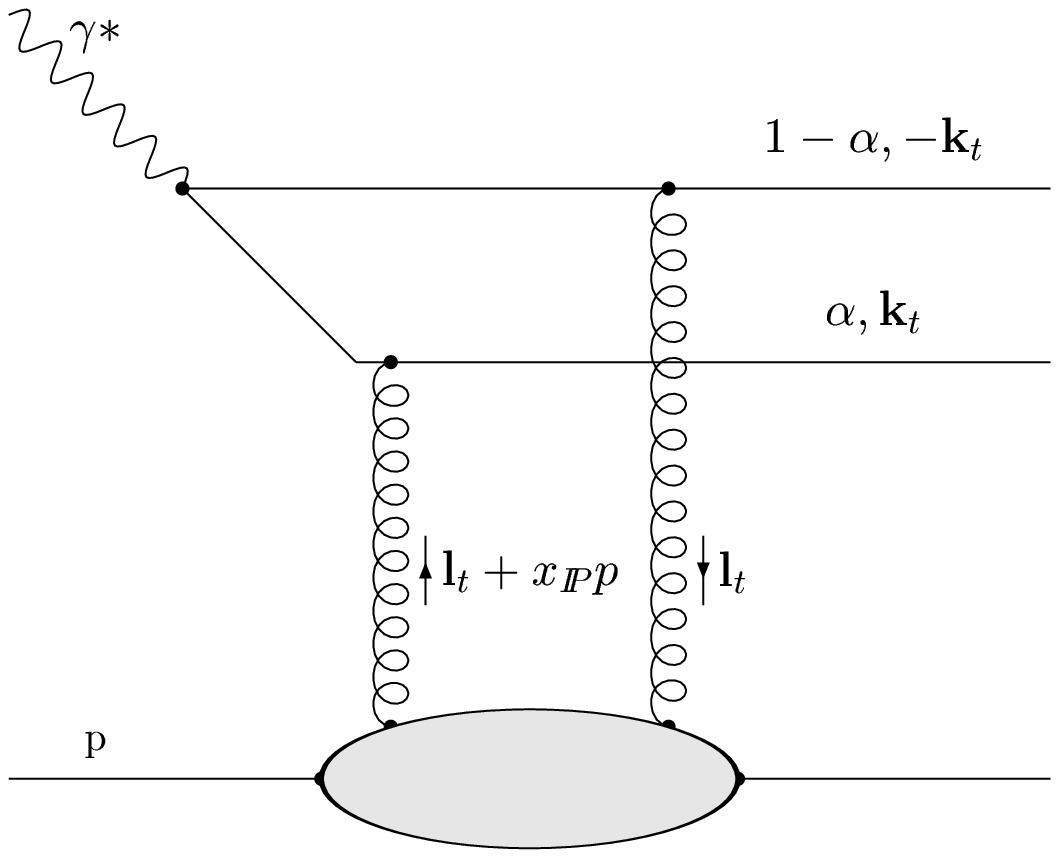}
  \includegraphics[height=6cm,width=6cm]{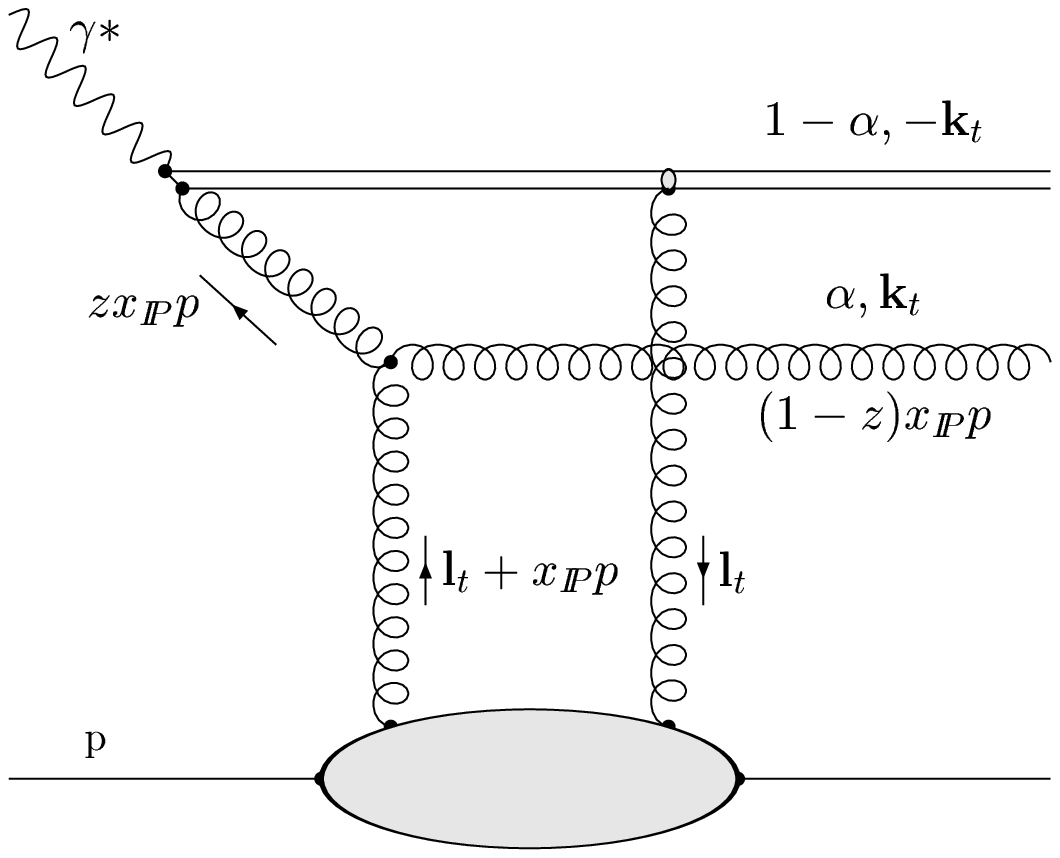}
  \caption{The q\={q} and q\={q}g contributions to $ \FD$.}
  \label{fig:feynman}
\end{figure}
\begin{figure}[htb]
\includegraphics[height=13cm,width=13cm]{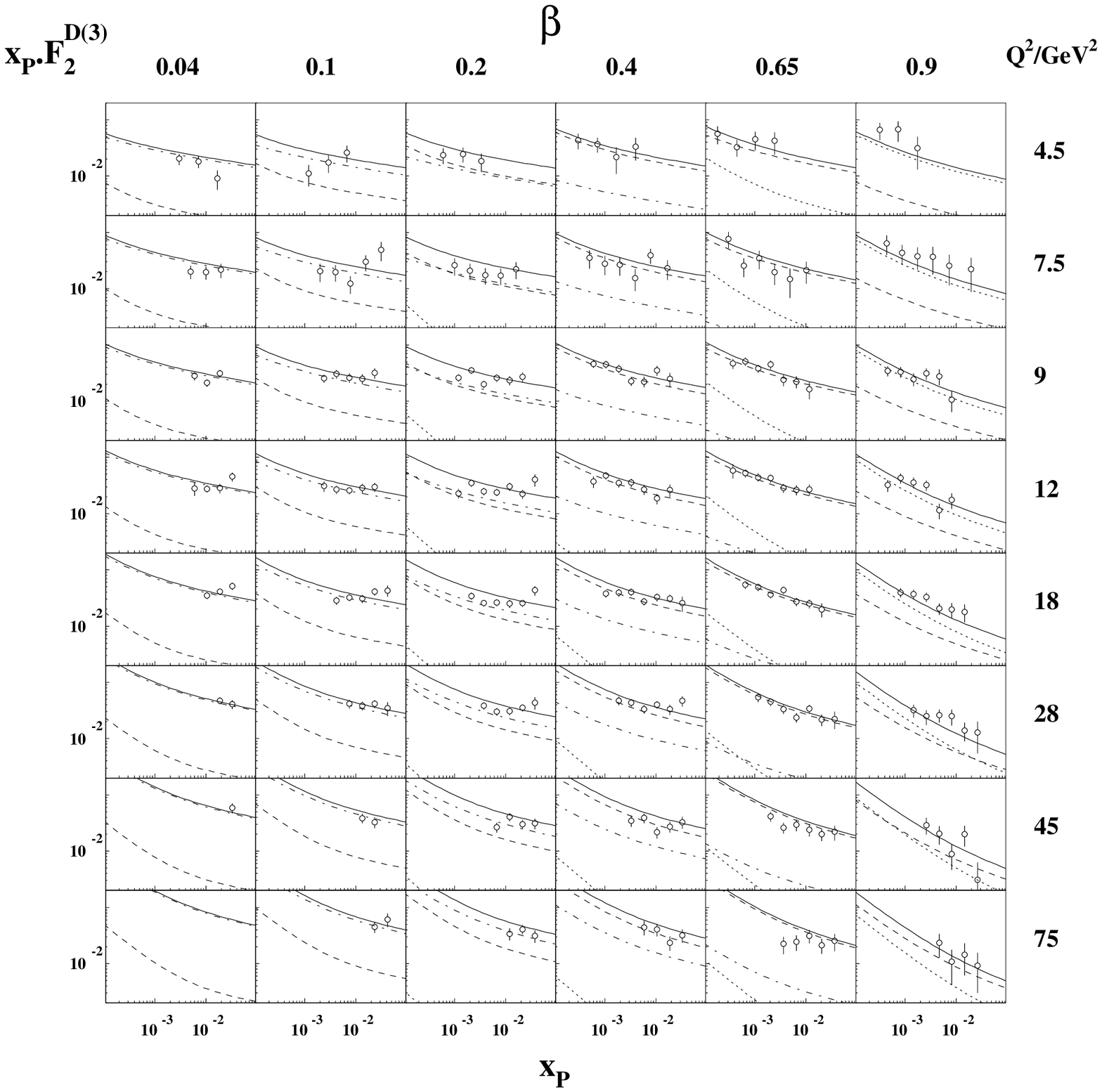}  
  \caption{Contributions to $\xp \FD$ compared with H1 1994 data. 
Full, dotted, dashed and dot dashed lines are the total, longitudinal q\={q}, transverse q\={q} and 
q\={q}g contributions respectively.}
  \label{fig:f2d3}
\end{figure}

To calculate the contribution of the quark antiquark dipole  to $ \FD$ we made use of expressions  derived from a momentum space treatment~\cite{bib:mw.der,bib:mw.kgb.diff}. Also, we calculated the contribution of the higher q\={q}g Fock state using an effective two gluon dipole description from the same source. Typical Feynman diagrams are shown in Figure~\ref{fig:feynman}. For compatibility with this approach, we must replace $s$ by $\tilde{s}$, where \(\tilde{s}= \qq(1/\xp - 1)\).
First defining
\begin{equation}
  \Phi_{0,1}  \equiv  \int_{0}^{\infty}r \D r K_{0 ,1}(\epsilon r)\sigma_{d}(r, \tilde{s})J_{0 ,1}(kr)
  \int_{0}^{\infty}r \D r f(r)K_{0 ,1}(\epsilon r)\sigma_{d}(r, \tilde{s})J_{0 ,1}(kr),  
\end{equation}
we have for the longitudinal and transverse q\={q} components respectively
\begin{equation}
  x_{I\!\!P}F^{D}_{q\bar{q},L}(Q^{2}, \beta, x_{I\!\!P})=
\frac{3 Q^{6}}{32 \pi^{4} \beta b}\cdot \sum_{f=1}^{3}e_{f}^{2} 
\cdot 2\int_{\alpha_{0}}^{1/2} \D \alpha \alpha^{3}(1-\alpha)^{3} \Phi_{0}
\end{equation}
\begin{equation}
 x_{I\!\!P}F^{D}_{q\bar{q},T}(Q^{2}, \beta, x_{I\!\!P}) =  
     \frac{3 Q^{4}}{128\pi^{4} \beta b} \cdot \sum_{f=1}^{3}e_{f}^{2} \cdot
 2\int_{\alpha_{0}}^{1/2} \D \alpha \alpha(1-\alpha) 
\left\{ \epsilon^{2}[\alpha^{2} + (1-\alpha)^{2}] \Phi_{1} + m^{2} \Phi_{0}  \right\}   
\end{equation}
where the lower limit of the integral over $\alpha$ is given by
\(
\alpha_{0} = (1/2)\left(1 - \sqrt{1 - 4m_{f}^{2}/M_{X}^{2}}\right)
\)
and $b$ is the slope parameter, which we have taken as 7.2~\cite{bib:slope}.
For the q\={q}g term we have 
\begin{eqnarray}
   \lefteqn{x_{I\!\!P}F^{D}_{q\bar{q}g}(Q^{2}, \beta, x_{I\!\!P}) 
  =  \frac{81 \beta \alpha_{S} }{512 \pi^{5} b} \sum_{f} e_{f}^{2} 
 \int_{\beta}^{1}\frac{\mbox{d}z}{(1 - z)^{3}} 
 \left[ \left(1- \frac{\beta}{z}\right)^{2} +  \left(\frac{\beta}{z}\right)^{2} \right] } \\
  & \times & \int_{0}^{(1-z)Q^{2}}\mbox{d} k_{t}^{2} \ln \left(\frac{(1-z)Q^{2}}{k_{t}^{2}}\right) 
   \left[ \int_{0}^{\infty} u \mbox{d}u \; \sigma_{d}(u / k_{t}, \tilde{s}) 
   K_{2}\left( \sqrt{\frac{z}{1-z} u^{2}}\right)  J_{2}(u) \right]^{2}
\end{eqnarray} with $\alpha_{S} = 0.2$.
(We have inserted a missing factor of 1/2 compared with the expression in~\cite{bib:mw.kgb.diff}.) 
This expression diverges if our parameterisation is used as it stands. However, this is due solely to the mild divergence  after \dcs\ saturates in $r$ of the factor
$(r^{2}s)^{\lambda_{S}}$ as $r \rightarrow \infty$. This behaviour at large $r$ is not determined by data and is an artefact of the parameterisation.
Hence  we have imposed a saturation value for \dcs\ of its value at $r$ = 2\ fm for all higher $r$.
Plots of the contributions to $\xp \FD$ calculated from these expressions are compared with H1 1994 data~\cite{bib:H1}
 in Figure~\ref{fig:f2d3}.
Agreement is good, even at low $\beta$ where the q\={q}g term dominates.  Comparison with ZEUS 1994 data~\cite{bib:ZEUS} also gives good agreement overall but with deviations at larger \qq\ values for small and moderate $\beta$.
\section{Conclusions}
We have successfully predicted the diffractive structure function \FD using a parameterised dipole cross section obtained from electro- and photoproduction data. Unlike the model proposed in~\cite{bib:mw.kgb.diff,bib:mw.kgb.fit}, the parameterisation exhibits effective saturation in $r$ only, with no saturation in the energy variable $s=W^{2}$. Agreement with data is reasonable, leading to the conclusion that the HERA data do not necessarily indicate such saturation at present energies.
\section{Acknowledgements}
GRK would like to thank PPARC for a Studentship.
This work was supported in part by the EU Fourth Programme `Training and 
Mobility of Researchers', Network `Quantum Chromodynamics and the Deep 
Structure of Elementary Particles', contract FMRX-CT98-0194 (DG 12-MIHT). 
\end{document}